\documentclass[superscriptaddress,showpacs,floatfix,preprintnumbers,amsmath,amssymb]{revtex4}

\usepackage{graphicx} 
\usepackage{dcolumn}  
\usepackage{subfigure}

\newcommand{\beq}{\begin{equation}}
\newcommand{\eeq}{\end{equation}}
\newcommand{\beqn}{\begin{eqnarray}}
\newcommand{\eeqn}{\end{eqnarray}}

\newcommand{\gevc}{\ensuremath{\,{\mathrm{GeV}/c^2}}}
\newcommand{\mevc}{\ensuremath{\,{\mathrm{MeV}/c^2}}}
\newcommand{\mev}{\ensuremath{\,{\mathrm{MeV}}}}

\newcommand{\Br}{\ensuremath{\mathcal{B}}}

\newcommand{\pp}{\ensuremath{\psi(2S)}}
\newcommand{\cho}{\ensuremath{\chi_{c1}}}

\newcommand{\ccr}{\ensuremath{(c\bar{c})_{\mathrm{res}}}}
\newcommand{\ccrp}{\ensuremath{\ccr\,\pi^+}}

\newcommand{\Dsj}{\ensuremath{D_{sJ}^-}}
\newcommand{\Ds}{\ensuremath{D_{s}^-}}
\newcommand{\Dsr}{\ensuremath{D_{s}(2S)^-}}
\newcommand{\Dssr}{\ensuremath{D_{s}(2S)^{*-}}}

\newcommand{\Daa}{\ensuremath{\bar{D}^{(*)}}}
\newcommand{\Dba}{\ensuremath{D^{(*)}}}
\newcommand{\Daba}{\ensuremath{\Daa\Dba}}

\newcommand{\Da}{\ensuremath{\bar{D}}}

\newcommand{\Dab}{\ensuremath{(D\bar{D}{}^*)^+}}

\newcommand{\Dst}{\ensuremath{D^{*}}}

\newcommand{\aB}{\ensuremath{\bar{B}}}

\newcommand{\Z}{\ensuremath{Z^+}}
\newcommand{\Za}{\ensuremath{``Z''}}
\newcommand{\Zp}{\ensuremath{Z(4430)^+}}
\newcommand{\Zo}{\ensuremath{Z_1(4050)^+}}
\newcommand{\Zt}{\ensuremath{Z_2(4250)^+}}

\newcommand{\ppp}{\ensuremath{\psi(2S)\pi^+}}
\newcommand{\chp}{\ensuremath{\chi_{c1}\pi^+}}

\newcommand{\ee}{\ensuremath{e^+ e^-}}

\newcommand{\tht}{\ensuremath{\theta}}
\newcommand{\thp}{\ensuremath{\theta^\prime}}
\newcommand{\thpp}{\ensuremath{\theta^{\prime\prime}}}

\begin{document}

\title{ \quad\\[0.5cm] \Large Charged charmonium-like states as
  rescattering effects in $\mathrm{ \aB \to D_{sJ}^- D^{(*)}}$ decays}

\author{P. Pakhlov} 

\address{Institute for Theoretical and Experimental Physics, Moscow,
  Russia}

\begin{abstract}

Using purely phenomenological approach we show that the peaking
structures observed in the \ppp\ and \chp\ mass spectra in $\aB \to
\pp\,(\cho) \pi^+ K$ decays can be result of $(D \bar{D}{}^{(*)})^+
\to \ccrp$ rescattering in the decays $\aB \to D_{sJ}^- (\to
\bar{D}{}^{(*)} K) D^{(*)}$. In particular, the position of the peak
in the chain $\aB \to \Dsr D^+ \to K^- \bar{D}{}^{*0} D^+ \to K^-
\ppp$ coincides well with the measured \Zp\ mass, assuming the mass of
\Dsr\ (the first radial excitation of \Ds) to be $2610\mevc$. The
widths of the \Zp\ peak is also well reproduced in this approach
independent on the width of \Dsr. Although the decay $\aB \to \Dsr
D^+$ has not been observed so far and even \Dsr-meson is not
discovered yet, this decay is expected to be large, and the mass of
\Dsr\ is predicted in the range $(2600-2650)\mevc$. The broad bump in
\chp\ spectrum can be attributed to the $\aB \to \Dssr D^+ \to K^-
\bar{D}{}^0 D^+$ decay observed with a large branching fraction
followed by rescattering $\bar{D}{}^0 D^+ \to \chp$.

\end{abstract}


\maketitle
\setcounter{footnote}{0}

The charmonium-like charged \Z\ states, seen by Belle in $\aB \to \ppp
K$ and $\aB \to \chp K$ decays remained puzzles for last few
years. Belle~\cite{Belle:z4430} observed the first \Zp\ state as a
sharp peak in \ppp\ mass spectrum near $M(\ppp) = 4430\mevc$ with
statistical significance of more than $6\,\sigma$. The main background
around the peak region is due to $\aB \to \pp K^*(892)$ and $\aB \to
\pp K_{2}^*(1430)$ decays. In the Belle analysis the two $K^{*(*)}$
states were vetoed to suppress this background. It was noted that
interference between different partial waves in the $\pi K$-system can
produce peaks, that are reflections of the $K^{*(*)}$
polarization. However, these effects should also produce additional
sharp structures nearby in $M(\ppp)$, which are not observed by
Belle. More detailed analysis was performed by
Belle~\cite{Belle:Dalitz} a year latter to prove quantitatively the
absence of fake peaks by the Dalitz fit over all signal events
including $K^{*(*)}$ regions. In this study Belle confirmed the
\Zp\ observation and found its parameters consistent with those
measured in the first paper. Although BaBar~\cite{BaBar:z4430} has
reported no evidence of \Zp\ in their $\aB \to \ppp K$ analysis, the
non-uniform structures are presented in their spectrum as well, and
both Belle and BaBar spectrum are consistent with each other.

Two broader peaks (\Zo\ and \Zt) were found by Belle~\cite{Belle:z12}
in the \chp\ mass spectrum in the analysis of $\aB \to \chp K^{0\-}$
decay. A Dalitz fit with a single resonance in the \Z\ channel is
favored over a fit with only $K^{*(*)}$-resonances and no \Z-fit by
more than $10 \,\sigma$. Moreover, a fit with two \Z\ resonances is
favored over the fit with only one resonance by $5.7 \,\sigma$.

\begin{figure}[htb]
\begin{center}
\begin{tabular}{cc}
\includegraphics[width=0.4\textwidth] {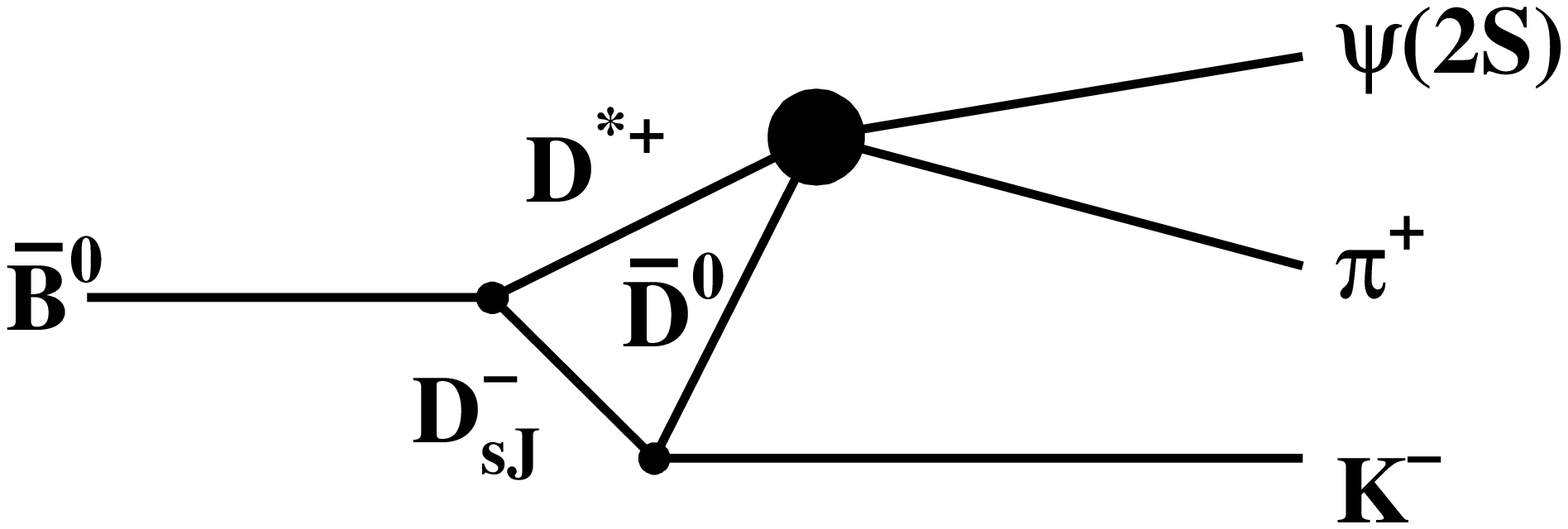} ~   &  ~  
\includegraphics[width=0.4\textwidth] {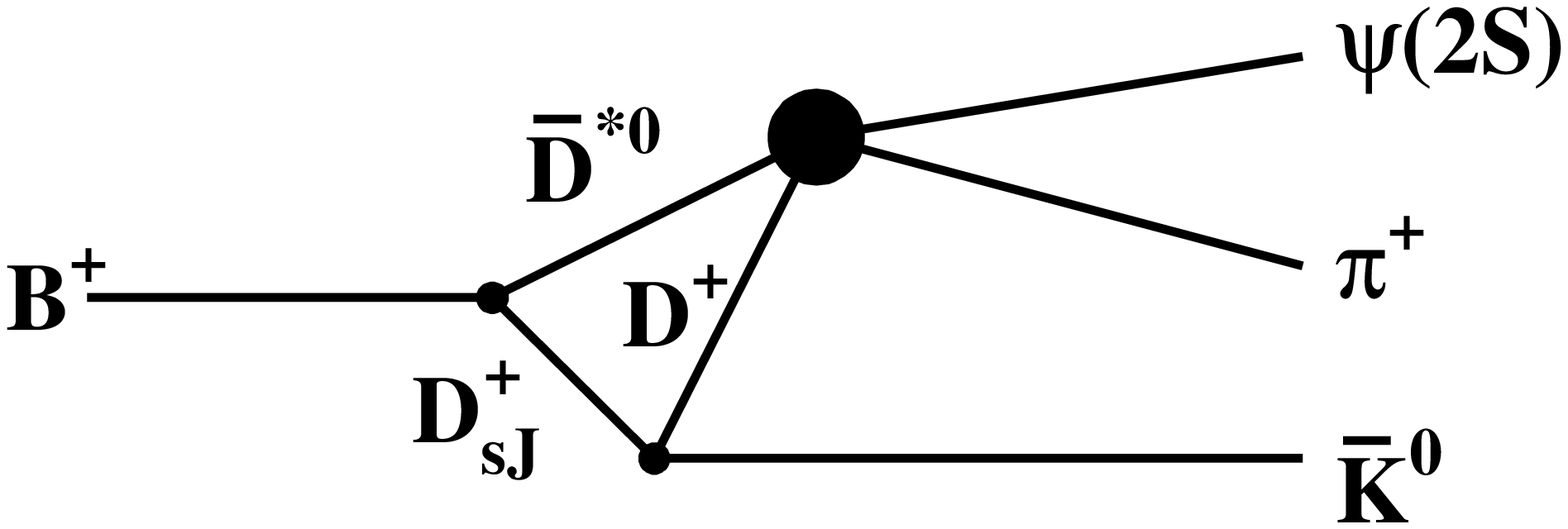} \\
\end{tabular}
\end{center}
\caption{Rescattering diagrams resulting in $\aB \to \ccrp K$ final
  state.}
\label{diag}
\end{figure}

If the observed \Z\ peaks are real states, they would necessarily be
exotic (non-conventional $q\bar{q}$) mesons, as their minimal
substructure consists of four quarks. Many attempts of explanation of
\Z's follow these observations including
molecular~\cite{molecule:z4430,molecule:z12},
tetraquark~\cite{tetraq}, hadrocharmonium states~\cite{hadroc} or cusp
effects~\cite{cusp}. In this Letter we demonstrate that the observed
peaks can be explained by the effect of rescattering in the decay
chain
\begin{equation}
\aB \to \Dsj  \Dba \quad  {\mathrm{followed ~~ by ~~ }} \Dsj \to \Daa K
\label{react}
\end{equation}
of \Daba-pair into charmonium+$\pi^+$, where one \Dba\ is directly
produced in \aB-decay, while \Daa\ is from the intermediate
\Dsj\ resonance (Fig.~\ref{diag}).

We assume that the \aB\ decay dynamics can be factorized from the
rescattering process. Under this assumption the mass of the
charmonium+$\pi^+$ combination is equal to those of \Daba. The mass
spectrum of \ccrp\ system produced via rescattering consists of
peaking structure(s) reflecting the \Dsj\ polarization in \aB-decay or
\Daba\ polarization in the formation of \ccrp. Here we denote the
$\Daba\equiv\ccrp$ as \Za\ and calculate \Za\ mass ignoring the
subsequent decay $\Za \to \ccrp$. However, angular momentum/parity
conservation impose restrictions on \Za\ production depending on the
final \ccrp\ state that should be taken into account. In particular,
\ppp\ and \chp\ systems have different spin-parity, thus different
decay chains of the type~(\ref{react}) should be proposed for the
explanation of the peaks in their spectra. We also assume that
rescattering of \Daba\ in $S$-wave dominates.

There are many decay chains~(\ref{react}) that can provide the
required conditions; all of them should be considered taking into
account interference between them. We are now limiting ourselves by
searching for the dominant contribution that can roughly reproduce the
features of the observed \ccrp\ spectra. We note that orbital
\Ds\ excitations are hardly suitable for our explanation: $j=1/2$
states are below $D^{(*)} K$ threshold, while \aB-decays into $j=3/2$
states are strongly suppressed ($\Br \lesssim
10^{-4}$~\cite{PDG}). Radial \Ds\ excitations could be better
candidates: the decay $\aB \to D_s(2700)^- D$ was observed by Belle
with relatively large branching fraction $\Br(\aB \to D_s(2700)^- D)
\times \Br( D_s(2700)^- \to \Da K) \sim
10^{-3}$~\cite{{Belle:BDs1}}. The quantum numbers of the $D_s(2700)^-$
have been measured to be $J^P=1^-$ and this state is likely to be a
radial excitation of $D_s^{*-}$ (\Dssr). If this is really true, the
pseudo scalar state, \Dsr, have a mass of $(2600-2650)\gevc$ (expected
$2S^1 - 2S^3$ splitting is $(60-100) \mevc$~\cite{theor}), and should
decay predominantly into $D^* K$. It is also expected that two body
\aB-decays into \Dsr\ are not suppressed.

We are first looking for an explanation of the \Zp\ peak. The
rescattering of $(D\bar{D})^+ \to \ppp$ is forbidden by parity/angular
momentum conservation, while the diagram $\Dab \to \ppp$ is allowed
for both initial and final systems in the $S$-wave. The two $\aB \to
D^+ \bar{D}{}^{*0} K$ and $\to \bar{D}{}^0 D^{*+} K$ decay modes that
include all intermediate states (as expected mostly \Dsj) has a
relatively large branching fraction of $\sim 10^{-2}$~\cite{PDG}.
\Dab\ system should form a pseudo-state with the spin equal to one and
positive parity ($J^P=1^+$), thus the parity of the intermediate $\Dsj
D^{(*)}$ combination should be also positive. The decay chain $\aB \to
\Dsr D^+$ followed by $\Dsr \to \bar{D}{}^{*0} K^-$ matches this
parity constraint. Another allowed decay chain that can result in
\ppp\ is $\aB \to \Dssr (\to \bar{D}^0 K) D^{*+}$ in case of $S$
($D$)-wave between \Dssr\ and $D^{*+}$. We calculate the matrix
elements of these decays in the helicity formalism
\begin{equation}
\mathcal{M}(M_{\Dab}) \sim \Big| ~ \sum_{\lambda_{D^*}}~ a_{\lambda_{D^*}} ~
A_{BW}(M_{\bar{D}{}^{(*)}K}) ~ ~ D_{D_{sJ}^-}(\tht,\lambda_{D^*}) ~ ~
D_{D^*}(\thp, \lambda_{D^*}) ~ ~ D_Z (\thpp) ~ \Big|^2~,
\end{equation}
where $A_{BW}$ is a Breit-Wigner function for a corresponding
\Dsj-resonance; $D(\theta,\lambda_{D^*})$ is it's angular part
depending on the \Dsj\ decay angle, \tht, and helicity of \Dst,
$\lambda_{D^*}$. (In case of the second decay chain the angular term
depends on the \Dssr\ helicity; however here we use equality
$\lambda_{D^*(2S)}=\lambda_{D^*}$). The next term, $D_{D^*}(\thp,
\lambda_{D^*})$, is responsible for the rotation of the \Dst\ spin
from the \Dsj\ (or \aB\ in case of the second decay chain) to
\Za\ rest frame. We note that in the \Za\ rest frame the
\Dst\ helicity is fixed to zero. Finally, $D_Z (\thpp)$ provides
formation of \Za\ from helicity-0 \Dst\ and \Da. We list explicitly
angular contributions to the matrix elements in the upper lines of
Table~\ref{tab:wigner} for two considered decays: $\aB \to \Dsr D$ and
$\aB \to \Dssr D^*$ in terms of (small) Wigner functions. In the later
decay we assume that $S$ (over $D$) wave dominates.
\begin{table}[]
\caption{The angular parts of the $\aB \to \Dsj \Dba \to (\Daba) K \to
  \Za K $ chain matrix element.}
\begin{center}
\begin{tabular}{l|c|c|c|c|c|c|c}\hline \hline

& $\aB$ decay mode & \Dsj\ decay mode & $\lambda_{D^*}$ &
  $a_{\lambda_{D^*}}$ & $D(\tht,\lambda_{D^*})$ & $D_{D^*}(\thp,
  \lambda_{D^*})$ & $D_Z (\thpp)$ \\ \hline \hline

& $\Dsr D$ & $\bar{D}{}^* K$ & 0 & 1 & 1 & $d^1_{0,~0}$ & $d^1_{0,~0}$
  \\ \cline{2-8}

\ppp & $\Dssr D^*$ & $\bar{D} K$ & $\pm 1$ & $\sqrt{1/3}$ &
$d^1_{0,~\pm 1}$ & $d^1_{\pm1,~0}$ & $d^1_{0,~0}$ \\

& $\Dssr D^*$ & $\bar{D} K$ & 0 & $- \sqrt{1/3}$ & $d^1_{0,~0}$ &
$d^1_{0,~0}$ & $d^1_{0,~0}$ \\ \hline \hline

\chp & $\Dssr D$ & $\bar{D} K$ & $-$ & 1 & $d^1_{0,~0}$ & 1 &
$d^1_{0,~0}$ \\ \hline \hline
\end{tabular}
\end{center}
\label{tab:wigner}
\end{table}

\begin{figure}
\begin{center}
\includegraphics[width=0.88\textwidth] {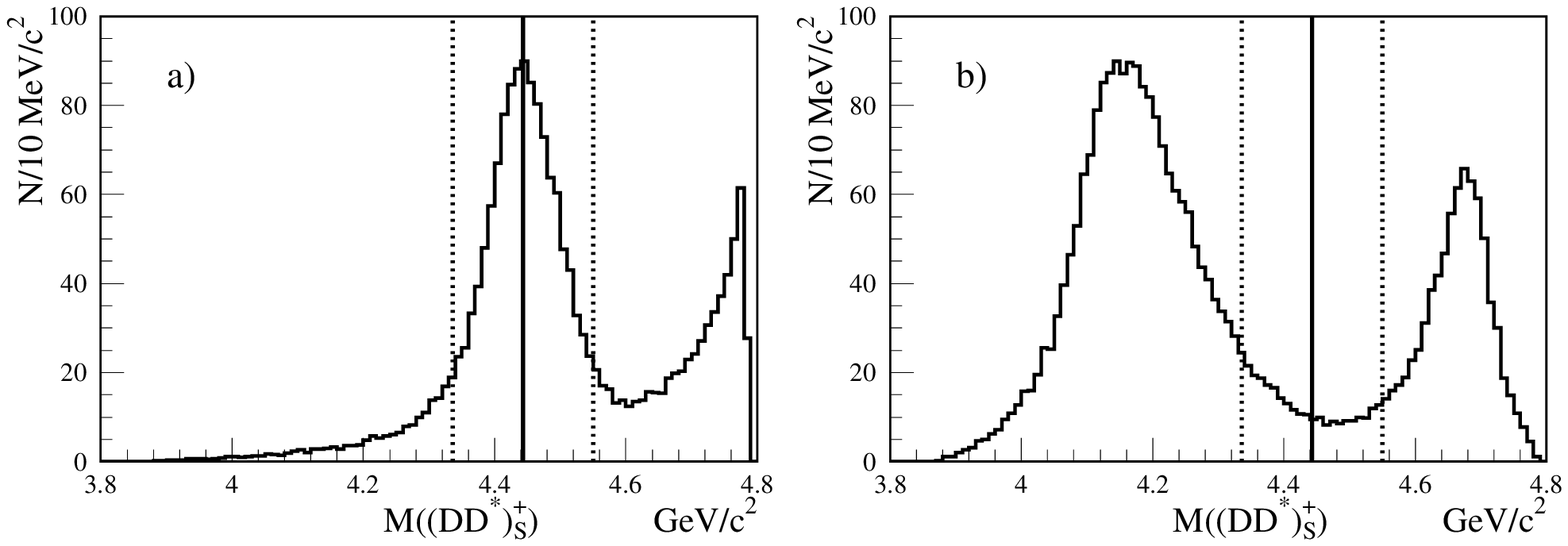} 
\begin{tabular}{cc}
~ \includegraphics[width=0.34\textwidth] {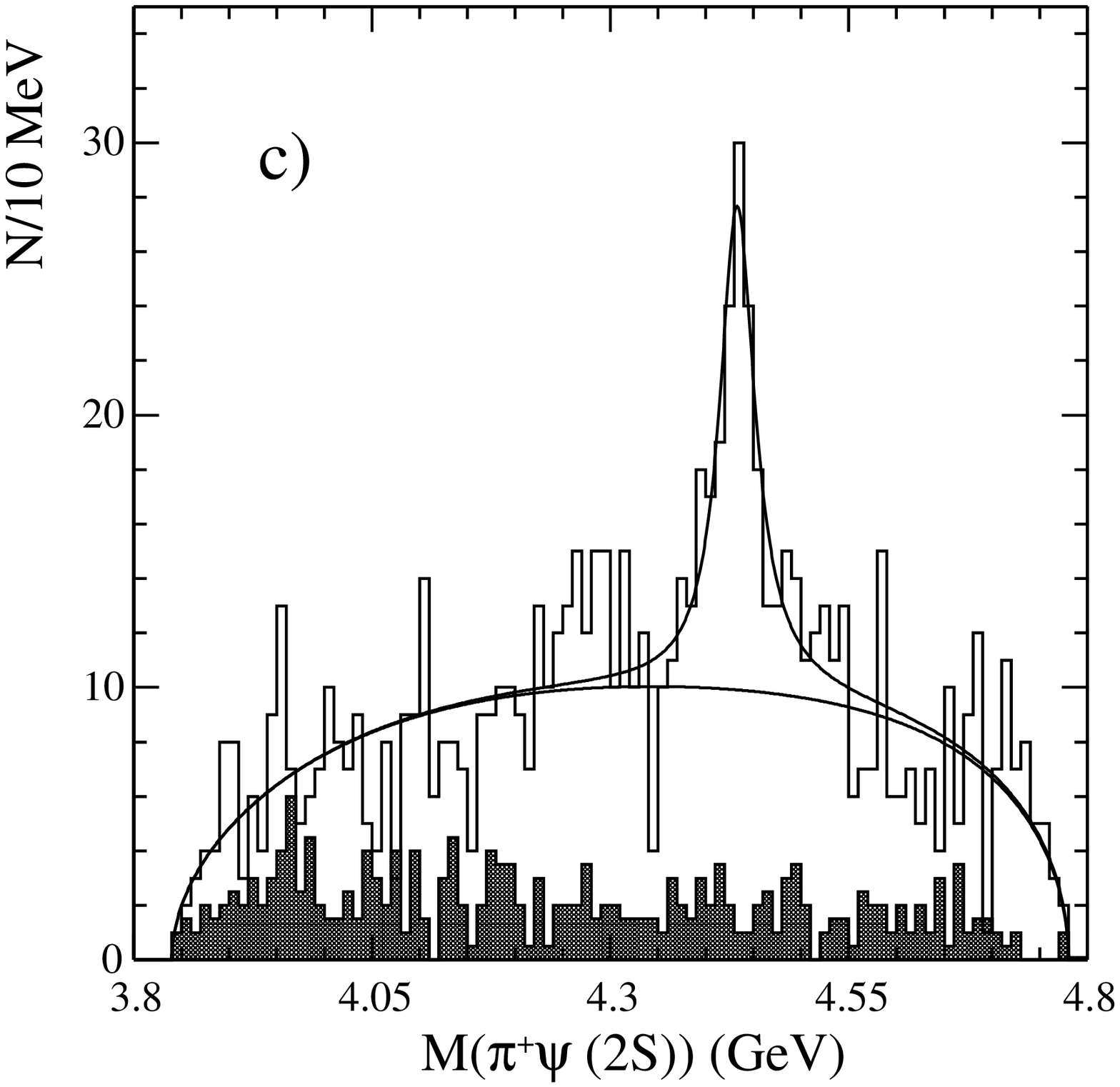} & \,
\includegraphics[width=0.5\textwidth] {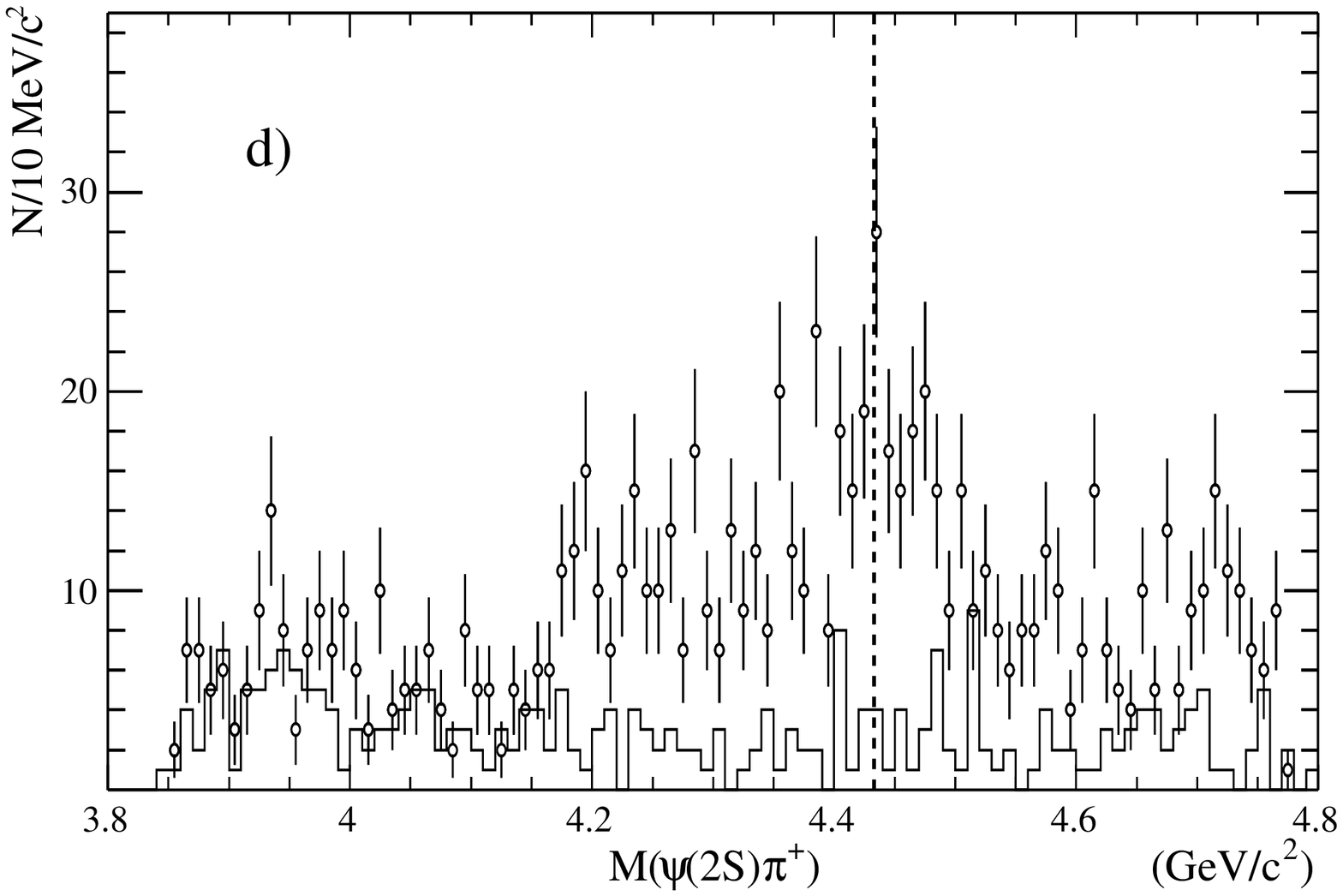} \\
\end{tabular}
\end{center}
\caption{The $M(\Za)$ spectra in the decays a) $\aB \to \Dsr D$; b)
  $\aB \to \Dssr \Dst$. The \ppp\ spectra in the selected $B \to \ppp
  K^{0\-}$ events in the c) Belle and d) BaBar data.}
\label{mpp}
\end{figure}

We calculate the $M(\Za)$ spectra using Monte Carlo simulation. In
this calculations we assume the mass of \Dsr\ to be $2610\mevc$ and
its widths to be $60\mev$; the \Dssr\ parameters are fixed to PDG
values~\cite{PDG}. The obtained spectra are presented for $\aB \to
\Dsr D$ and $\aB \to \Dssr \Dst$ decays in Fig.~\ref{mpp} a) and b),
respectively. 
The solid lines show the $\Zp$ peak position; the dashed
lines show $\pm \Gamma_{Z^+}$ window (as determined in
Ref.~\cite{Belle:Dalitz}). The main peak position and width in
Fig.~\ref{mpp} a) are close to the experimentally measured values for
\Zp. The contribution from $\aB \to \Dssr \Dst$ (Fig.~\ref{mpp} b) may
be responsible for another broad excess of signal events over
combinatorial background in the region above 4.1\gevc, that can be
seen in both Belle and BaBar data. For comparison the experimental
spectra seen at Belle and BaBar are shown in Fig.~\ref{mpp} c) and d),
respectively. We note that the instrumental reconstruction efficiency
in the region of $(4.6-4.75)\gevc$ drops very sharply, as this mass
region corresponds to a low center-of-mass momentum of kaon. Thus
experimental bias can hide the accompanying high mass structure which
can be seen in our Monte Carlo spectra.

\begin{figure}[htb]
\begin{tabular}{cc}
\includegraphics[width=0.44\textwidth] {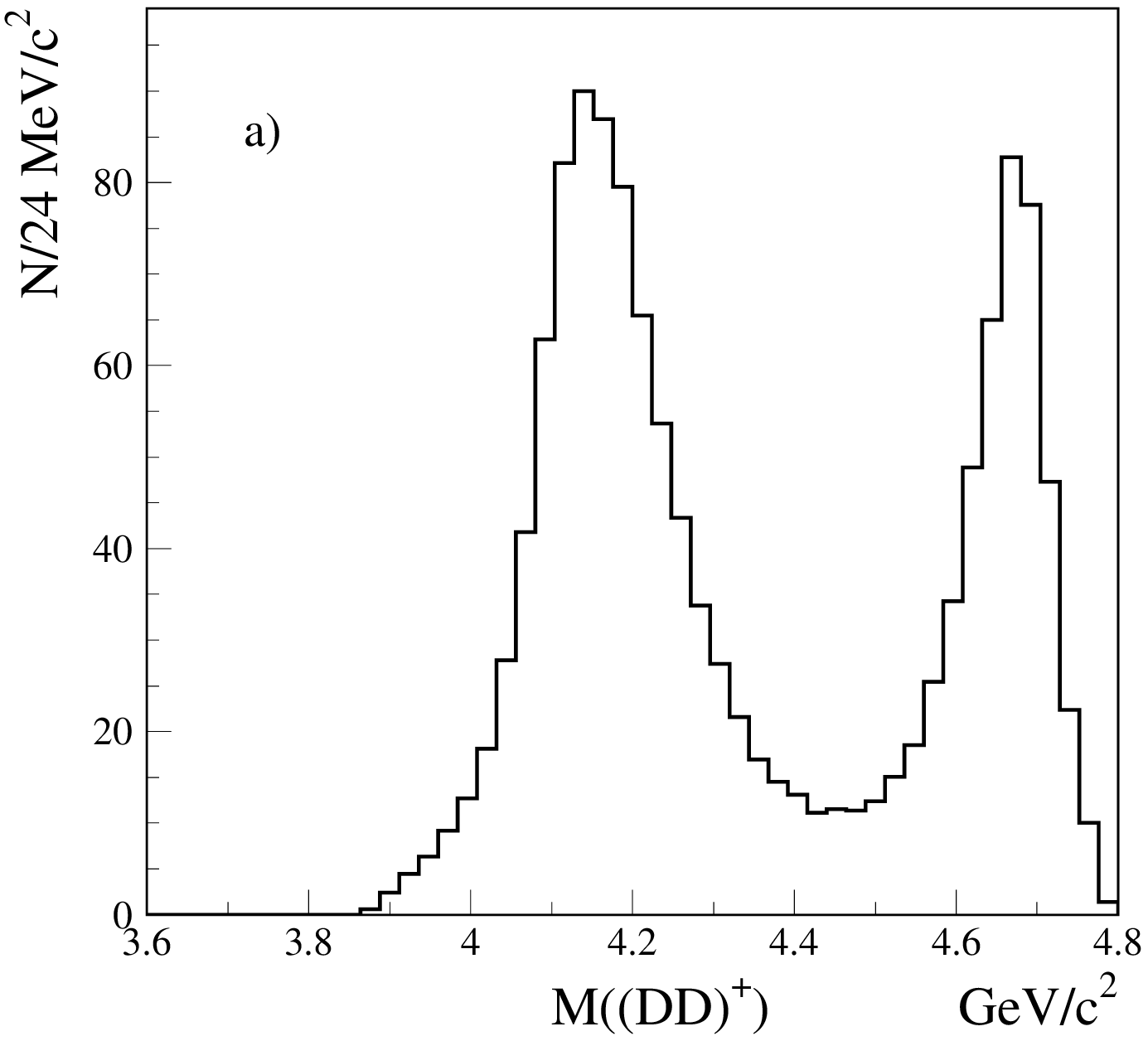} ~ & ~ 
\includegraphics[width=0.417\textwidth] {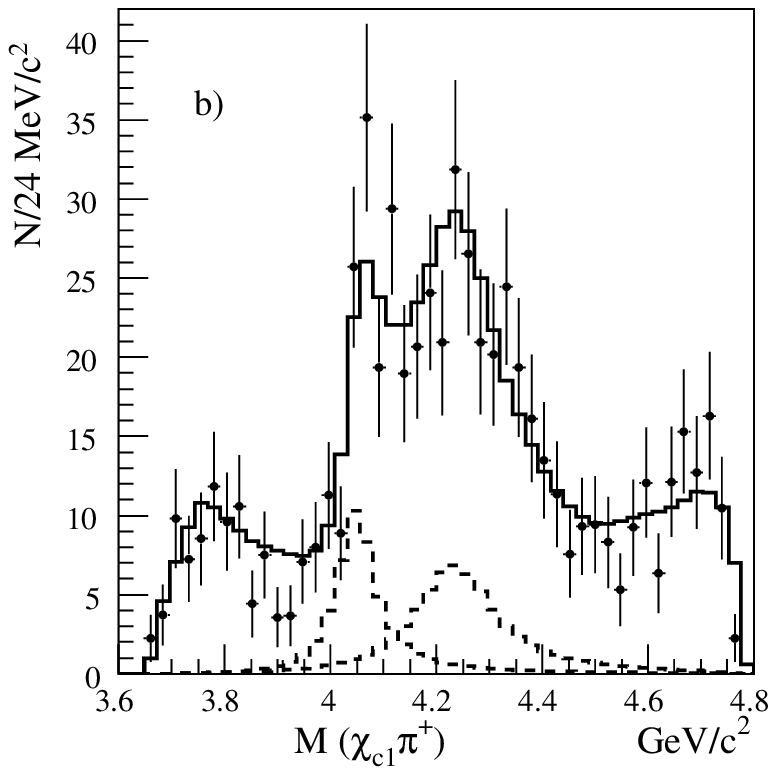} \\
\end{tabular}
\caption{a) The $M(\bar{D} D)$ spectrum in the decay $\aB \to \Dssr
  D$. b) The $M(\chp)$ spectrum in the Belle data.}
\label{mdd}
\end{figure}

For the explanation of the broad bump in \chp\ spectra around
$4-4.4\gevc$ region another chain of the type~(\ref{react}) with
negative parity of the final state has to be proposed. At least one
$P$-wave is required in the rescattering $(D^{(*)} \bar{D}{}^{(*)})^+
\to \chp$ to provide parity conservation. The simplest one
$(D\bar{D})^+_S \to (\chp)_P$ can be provided by the known decay $\aB
\to \Dssr D$. The Monte Carlo $(D\bar{D})^+$ mass spectrum from this
decay is shown in Fig.~\ref{mdd} a). The angular part was generated
according to the elements listed in Table~\ref{tab:wigner}. For
comparisons the Belle \chp\ spectrum from $\aB \to \chp K$ decays
(from Ref.~\cite{Belle:z12}) is shown in Fig.~\ref{mdd} b). We remind
that the experimental efficiency around high mass bump is
significantly lower, which can explain much smaller peak in the
experimental spectrum at $(4.6-4.8)\gevc$.

In summary, we show that the $\Dab \to \ppp$ rescattering in the decay
chain $\aB \to \Dsr D$, $\Dsr \to \bar{D}^* K$ can explain appearance
of a peaking structure in \ppp\ mass spectrum in $B \to \ppp K$ decays
around $4.43\gevc$. The broad structure in \chp\ mass spectrum around
$4-4.4\gevc$ can be also explained within the similar approach using
another decay chain $\aB \to \Dssr D$, $\Dssr \to \bar{D} K$ with
subsequent rescattering $(D\bar{D})^+ \to \chp$. We note, that if our
explanation is valid for charged charmonium-like peaks observed in
\aB-decays, the similar mechanism should also reveal itself in other
processes. For example, large $\ee \to \ppp \pi^-$ cross section can
be also explained by rescattering from the $\ee \to D \bar{D}{}^{*}
\pi$ process followed by $D \bar{D}{}^{*} \to \pp \pi$. The later
cross section have been measured by Belle~\cite{Belle:ddspi}, though
with large uncertainty. The peaks in the former cross section
(considered now as the $Y(4350)$ and $Y(4660)$ states) may be related
to the interference of $\psi$-resonances and many intermediate states
({\it e.g.} $D_1 \bar{D}$, $D_2 \bar{D}$, $D_2 \bar{D}^*$, etc.) in
their decays with different $D^*$ helicities resulting in the $D
\bar{D}{}^* \pi$ final state. In particular, large $\ee \to \psi(4415)
\to D D_2$ cross section~\cite{Belle:psi} can serve as a hint of the
vicinity of the $Y(4350)$ and $Y(4660)$ states to the $\psi(4415)$
resonance. However, the quantitative calculations are still beyond our
capabilities mainly because of lack of experimental data.

The author is very grateful to A. Bondar, K. Chilikin and T. Uglov for
useful discussions. We acknowledge support of the Russian State Atomic
Energy Corporation ``Rosatom'' under contract No. H.4e.45.90.10.1078.

\end{document}